\documentclass[reprint,amsmath,amssymb,aps,prl]{revtex4-2}

\usepackage{physics}
\usepackage{graphicx}
\usepackage{dcolumn}
\usepackage{bm}
\usepackage{color,sidecap,colortbl}

\begin{document}


\title{Transport controlled by Poincar\'e orbit topology in a driven inhomogeneous lattice gas}

\author{Alec Cao} 
\author{Roshan Sajjad} 
\author{Ethan Q.\ Simmons} 
\author{Cora J.\ Fujiwara}
\author{Toshihiko~Shimasaki}
\author{David M.\ Weld}
\affiliation{Department of Physics, University of California, Santa Barbara, California 93106, USA}
 
\begin{abstract}
In periodic quantum systems which are both homogeneously tilted and driven, the interplay between drive and Bloch oscillations controls transport dynamics.  Using a quantum gas in a modulated optical lattice, we show experimentally that inhomogeneity of the applied force leads to a rich new variety of dynamical behaviors controlled by the drive phase, from  self-parametrically-modulated Bloch epicycles to adaptive driving of transport against a force gradient to modulation-enhanced monopole modes.
Matching experimental observations to fit-parameter-free numerical predictions of time-dependent band theory, we show that these phenomena can be quantitatively understood as manifestations of an underlying inhomogeneity-induced phase space structure, in which topological classification of stroboscopic Poincar\'e orbits controls the transport dynamics.
\end{abstract}

\maketitle

Spatially periodic quantum systems exhibit an oscillatory response to static forces~\cite{bloch,zener}. Any applied modulation can interact with Bloch oscillations, resulting in phenomena ranging from super-Bloch dynamics~\cite{sbo_nagerl} to high-harmonic generation~\cite{GhimireReisHHGreview}.  In this work, we experimentally explore the consequences of breaking the position-independent character of Bloch oscillations with an inhomogeneous field, which qualitatively transforms the phase space structure of the system and generates an array of new transport phenomena. The recently-observed position-space character of Bloch oscillations~\cite{PSBO-PRL} plays a key role, admixing an intrinsic self-parametric modulation to all Bloch oscillators.

The experiments we describe use a quantum gas in an optical lattice. Cold-atom experiments have long provided a flexible platform for exploring Bloch oscillations and related fundamental features of transport in crystals~\cite{salomon-blochoscs,kasevichatomintf, nagerlinteractionBOs, sbo_nagerl,Preiss1229,AMlatts-Tino,PSBO-PRL}. Modulated effective electric fields have been used to investigate Wannier-Stark ladder resonances~\cite{wsresonance_raizen}, modulation-assisted tunneling~\cite{delocalization_italians,pat_italians}, coherent spatial mode manipulation~\cite{transport_italians}, and super Bloch oscillations~\cite{sbo_nagerl}, complementing related theoretical studies~\cite{ogtheory_kenkre,wsmodulationtheory_french,tbalgebra_mossman,dynamiclocalization_holthaus,interactingtilt_kolovsky,sbotheory_monteiro}, and parametric lattice modulation has been applied to the study of quantum ratchet behavior~\cite{ratchet_weitz} and large Floquet-Bloch oscillations in hybridized bands~\cite{warpdrivePRL}.

\begin{figure}[hbt!]
    \centering
    \includegraphics[width = \columnwidth]{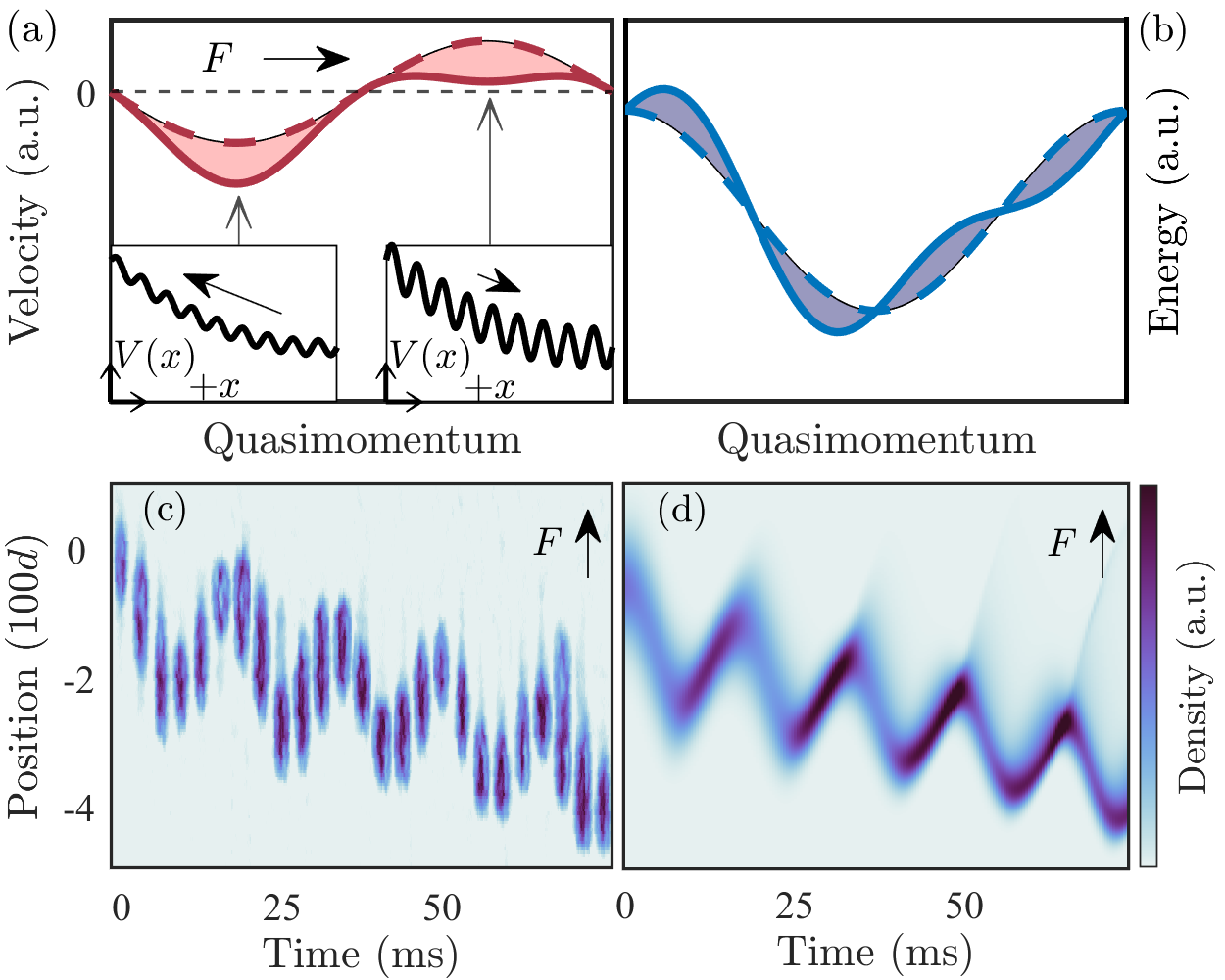}
    \caption{Transport in a modulated lattice. (a) Comparison between convective (solid) and static (dashed) group velocity for a Bloch-oscillating ensemble with resonantly-modulated tunneling. Shaded area indicates net spatial motion over a cycle. Insets show exaggerated real-space potential, with tunneling indicated by the length and direction of the arrow.
    (b) Corresponding convective (solid) and static (dashed) energy bands. (c) Ordered sequence of absorption images demonstrating transport against an applied force by a chirped adaptive drive (details in text).   (d) Theoretically predicted density evolution under the same conditions as (c).}
    \label{fig:setup}
\end{figure}

An overview of the experimental context of driven-lattice transport appears in Fig.~\ref{fig:setup}: modulating the lattice depth near the Bloch frequency gives rise to an asymmetric parametrically-varying ``convective'' group velocity and net transport during a Bloch cycle. The experiments begin with an optically-trapped Bose-Einstein condensate (BEC) of $10^5$ $^7$Li atoms adiabatically loaded into a 1D optical lattice with lattice spacing $d = 532$ nm, laser wave vector $k_L = \pi/d$, and recoil energy $E_R = \hbar^2 k_L^2 / 2m$, with $m$ the mass of $^7$Li. Interatomic interactions are set to zero by Feshbach tuning. The condensate starts in the crossed optical dipole trap at a position away from the center of a harmonic potential created by external electromagnets, so that when the dipole trap beams are abruptly turned off the atoms feel an inhomogeneous force and begin Bloch oscillating. All comparisons with theory are based on a Gaussian ensemble of spatial width $\sigma_x = 50d$ and momentum width $\sigma_k = .1k_L$; this non-Heisenberg-limited $\sigma_k$ is associated with the BEC experiencing inhomogeneous axial forces and consequent momentum broadening during the adiabatic lattice load. At $t_0 =9.3$ ms the BEC  reaches the edge of the Brillouin zone and we begin sinusoidal modulation of the lattice beam intensity.  Following the removal of the optical dipole trap, the system is described by the Hamiltonian
\begin{align}
    H = \frac{p^2}{2m} + \frac{V(t)}{2} \cos(2 k_L x) + \frac{1}{2}m \omega^2 x^2 - F x.
\label{eq:hamiltonian}
\end{align}
The magnetic trap frequency is \mbox{$\omega = 2\pi\times$ 15.5 Hz}, with initial local force $F = h/T_B d$ and Bloch period $T_B = 16.75$ ms. The time-varying lattice depth $V(t)$ is 
\begin{align}
    V(t) = \begin{cases}  V_0 \left[1+\alpha \sin(\varphi)\right], & -t_0 < t <0 \\ V_0 \left[1+\alpha \sin(\omega_D t+ \varphi)\right],&  \quad \quad t\geq0.\end{cases}
\label{eq:potential}
\end{align}
For all runs, the drive frequency is $\omega_D = 2 \pi \times $ 53.56 Hz; in the case of chirped driving, this is the frequency at $t=0$. The modulation depth is held at $\alpha = 0.24$ with average lattice depth $V_0 = 4.3 \, E_R$. Due to the drive and the force inhomogeneity, the angular Bloch frequency $\omega_B = 2 \pi / T_B$, drive period $T_D = 2 \pi / \omega_D$, and tunneling $J$ are all potentially time-varying quantities.  The critical parameter we manipulate to drive different dynamical behaviors is the initial drive phase $\varphi$.  

\begin{figure}[t]
    \centering
    \includegraphics[width = \columnwidth]{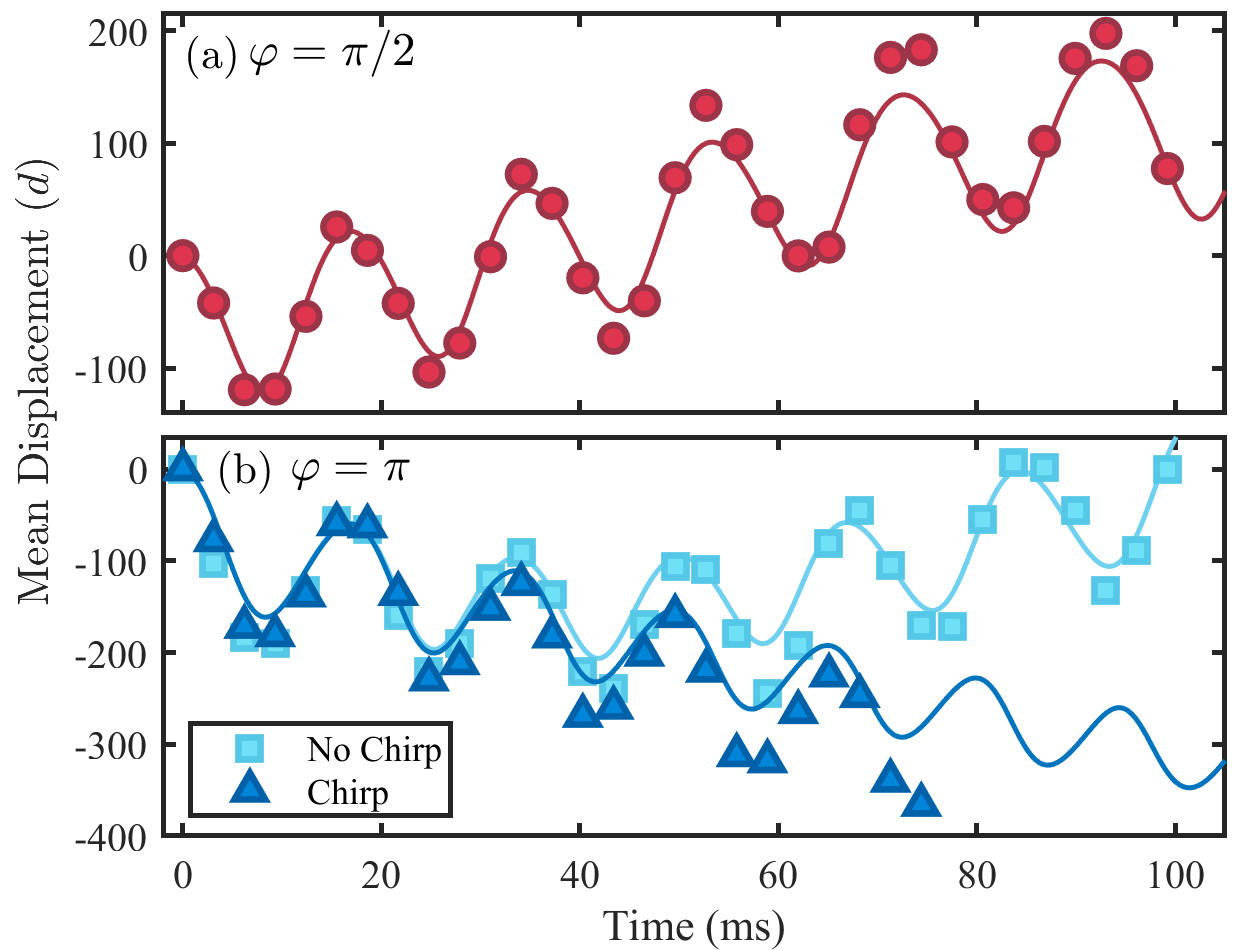}
    \caption{Directed Floquet-Bloch transport in an inhomogeneous force field. (a) Measured (points) and numerically predicted (line) mean atomic position as a function of time for a drive phase $\varphi=\pi/2$. The applied force points towards larger positive displacements. (b) Similar measurements (points) and numerical theory (lines) for an initial drive phase $\varphi=\pi$. A time-invariant drive frequency yields epicyclic motion due to force inhomogeneity (squares). An adaptive drive using a frequency chirp suppresses this behavior and extends the range of transport (triangles, and images in Fig.~\ref{fig:setup}c). }
    \label{fig:firstmoment}
\end{figure}

Fig. \ref{fig:firstmoment} shows data taken for drive phases experimentally found to be optimal for long-range pumping of the BEC both with and against the applied force. The results demonstrate pumping of the BEC over 200 lattice sites in just 5 Bloch cycles; the large increase in magnitude of transport rate as compared to Ref.~\cite{sbo_nagerl} can be attributed mainly to the low mass of $^7$Li. We observe optimal pumping along the direction of applied force at $\varphi=\pi/2$; as discussed below, this disagrees starkly with a theoretical description based on a homogeneous effective electric field, which predicts optimal transport along the direction of force for $\varphi=0$ and optimal transport against the force for $\varphi=\pi$. 

The observed dynamics are highly asymmetric in drive phase. Modulating at $\varphi = 3\pi/2$, exactly out of phase with the experimentally observed optimal condition for force-aligned pumping, does not produce directed transport. In experiments with a modulation phase of $\varphi=\pi$, the cycle-averaged velocity actually changes sign, as shown in Fig.~\ref{fig:firstmoment}b. This phenomenon is similar to super Bloch oscillations, though here the evolving relative phase between drive and Bloch oscillation does not result from a static detuning, but instead from a self-parametric modulation due to the spatial variation in Bloch frequency. Put differently, the force inhomogeneity eliminates the possibility of a global Wannier-Stark resonance, giving rise to slow oscillatory transport as a natural dynamical mode. 

 An adaptive driving protocol can recover directed monotonic pumping against an applied force even without a true Wannier-Stark resonance. Figs.~\ref{fig:firstmoment}b and 1c show experimental measurements of transport produced by an adaptive drive which includes a chirped drive frequency.  Intuitively, the chirp can be understood as stroboscopically maintaining the local Wannier-Stark resonance condition for a set of unevenly spaced ladders, or alternatively as optimizing the cycle-averaged spatial transport sketched in Fig.~1a by accounting for the average change in $\omega_B$ per cycle. These data were taken with a chirp rate of 115 Hz/s, causing an increase of $\omega_D$  by $2\pi \times$2.15 Hz each drive cycle. While here a linear chirp is shown to be effective for a linearly-varying force, the results suggest that higher-order, non-monotonic and piece-wise adaptive driving protocols could serve as flexible tools for engineering transport in arbitrary force landscapes. 

\begin{figure*}
    \centering
    \includegraphics[width = \textwidth]{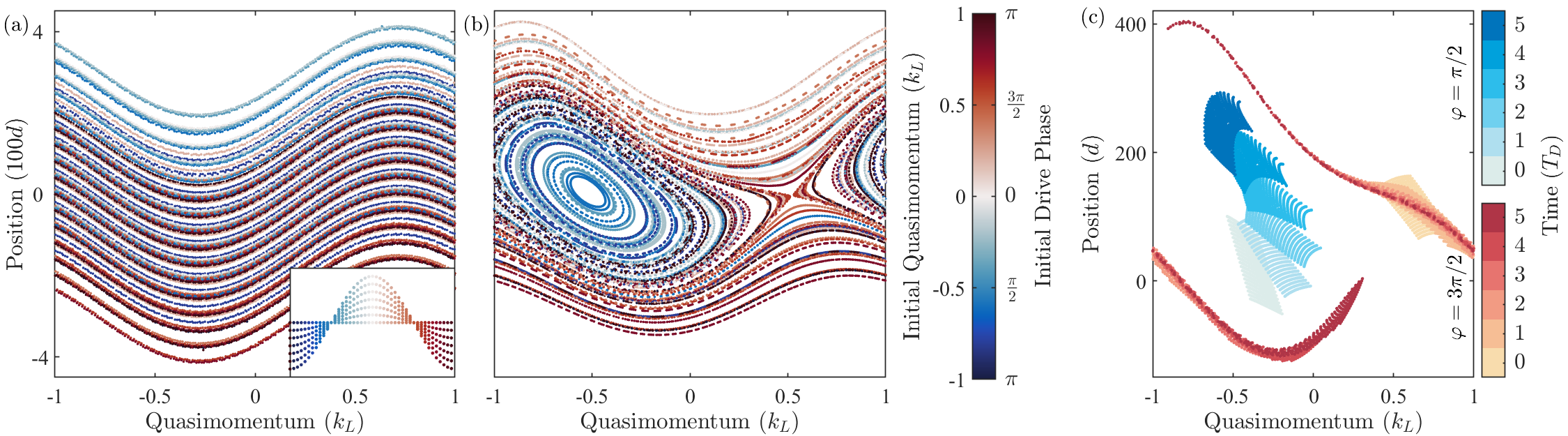}
    \caption{Effect of the inhomogeneity-induced phase space structure on transport. (a) Stroboscopic Poincar\'e map for a homogeneous force at 6~Hz detuning between drive and Bloch oscillation, showing super-Bloch-like oscillations wrapping the Brillouin zone. For panels (a) and (b) an 11$\times$11 grid of initial conditions spanning 400 lattice sites and the whole Brillouin zone was numerically evolved and plotted stroboscopically out to 200$\,T_D$ (longer than our longest experiment times). Colorbar indicates drive phase $\varphi$, or, equivalently, initial quasimomentum. Inset shows the stroboscopic Poincar\'e map on identical axes out to 6$\,T_D$ for zero detuning, for a single initial position  over the whole quasimomentum range; trajectories unwrap yielding linear vertical transport and two invariant quasimomenta. (b) Stroboscopic Poincar\'e map at the same detuning, for an inhomogeneous force matching our experimental conditions. Note the emergence of nontrivial fixpoints and topologically distinct classes of orbits. (c) Short-time portrait of the evolution of an ensemble for $\varphi= \pi/2$ (blue time) and $3\pi/2$ (red time), yielding stable transport and rapid spreading respectively as a result of the different fixpoint characteristics. The plotted sample is a 21$\times$21 grid spanning 1-$\sigma$ in both position and momentum. 
    The time colorbars match the definition of $t$ in in Eq. \ref{eq:hamiltonian} after adding .25 (blue) and .75 $T_D$ (red).}
    \label{fig:phasemap}
\end{figure*}

Next we discuss a simple analytic model of directed transport for a homogeneous force; this provides a useful framework for highlighting and understanding the qualitatively new phenomena introduced by force inhomogeneity. We consider a tight-binding model in the single-band approximation. For a sufficiently low-frequency drive we can define a time-dependent ground band dispersion $E(k,t) = -2J[V(t)] \cos(kd)$, with tunneling $J$ a function of the time-dependent lattice depth $V$, and $k$ the quasimomentum. In the semiclassical picture, the BEC moves at the group velocity
\begin{align}
    v_g(t) = \frac{2J[V(t)]d}{\hbar} \sin\left[k(t)d\right],
\label{eq:groupvelocity}
\end{align}
where $k(t)$ denotes the time-dependent quasimomentum. For weak modulation, $J$ varies to first order in time as $J[V(t)] \approx J(V_0) \left[1-\alpha_0 \sin(\omega_D t + \varphi)\right]$, with scaled modulation index $\alpha_0 = \alpha \abs{J'(V_0)} V_0/J(V_0)$; the prime indicates partial differentiation with respect to lattice depth, and for our experimental parameters $\alpha_0 \approx 1.15 \alpha$ as computed using Mathieu parameter relations for the band edges~\cite{holthaus-floquetengineering}. Under this approximation, the cycle-averaged spatial transport is
\begin{align}
    \Delta x \approx- \frac{2 \alpha_0 J(V_0) d}{\hbar} \!\int_0^{T_D}\!\!\!\!\! \sin(\omega_D t + \varphi) \sin(\int_0^{t} \omega_B(t') dt') dt. 
\label{eq:transport}
\end{align}
For a homogeneous force, $\omega_B(t)$ is a constant, and thus the quasimomentum evolves as $k(t) = (\omega_B t + \pi)/d$. The $\pi$ offset is introduced to match our experimental protocol. For resonant driving $\omega_D = \omega_B$, the average velocity is
\begin{align}
    \overline{v}_g = \frac{\alpha_0 v_0}{2} \cos(\varphi).
    \label{eq:meanvg}
\end{align}
On average the atoms travel at half the characteristic velocity $v_0 = 2 J(V_0) d/\hbar$ scaled by the modulation index $\alpha_0$ and the alignment between drive and Bloch cycles $\cos(\varphi)$. It is clear that $\varphi=0$ and $\varphi = \pi$ yield maximum pumping down and up the potential respectively. 

For small detuning between drive and Bloch frequencies, this homogeneous model predicts transport over many periods resulting from the effective evolution of $\varphi$. It is useful to compare and contrast these parametrically-driven dynamics to super Bloch oscillations~\cite{sbo_nagerl}: both can be schematically understood as resulting from modulation of the Wannier-Stark length $l=2J/F$, either by modulation of the numerator (this work) or the denominator (Ref.~\cite{sbo_nagerl}). In the language of nonlinear dynamics, the distinction is between forced and parametrically-excited oscillators, and our experiment is a quantum mechanical analogue of a parametrically-excited pendulum~\cite{parametricpendulum1,parametricpendulum2,parametricpendulum3}, with angle mapped to quasimomentum and angular momentum to position, in the regime of purely rotating solutions. 

While the intuitive mechanism for Floquet-pumped transport in an {\em inhomogeneous} force field can still be conceptually understood with this homogeneous framework, our measurements deviate qualitatively from these predictions as $\omega_B$ acquires parametric time dependence from the time-varying position. The  observed optimal phase for force-directed pumping (Fig.~\ref{fig:firstmoment}a) is in clear disagreement with the constant-force prediction of $\varphi=0$; in fact, Eq.~\ref{eq:meanvg} predicts a mean velocity of zero for $\varphi=\pi/2$. The effect of force inhomogeneity is even more pronounced when attempting Floquet pumping against the potential gradient: as shown in Fig.~\ref{fig:firstmoment}b, we observe a rapid change of the transport direction and no symmetry between opposite-phase drives.
These qualitative discrepancies are due mainly to the failure of the constant Bloch frequency approximation.

\begin{figure*}
    \centering
    \includegraphics[width = .97 \textwidth]{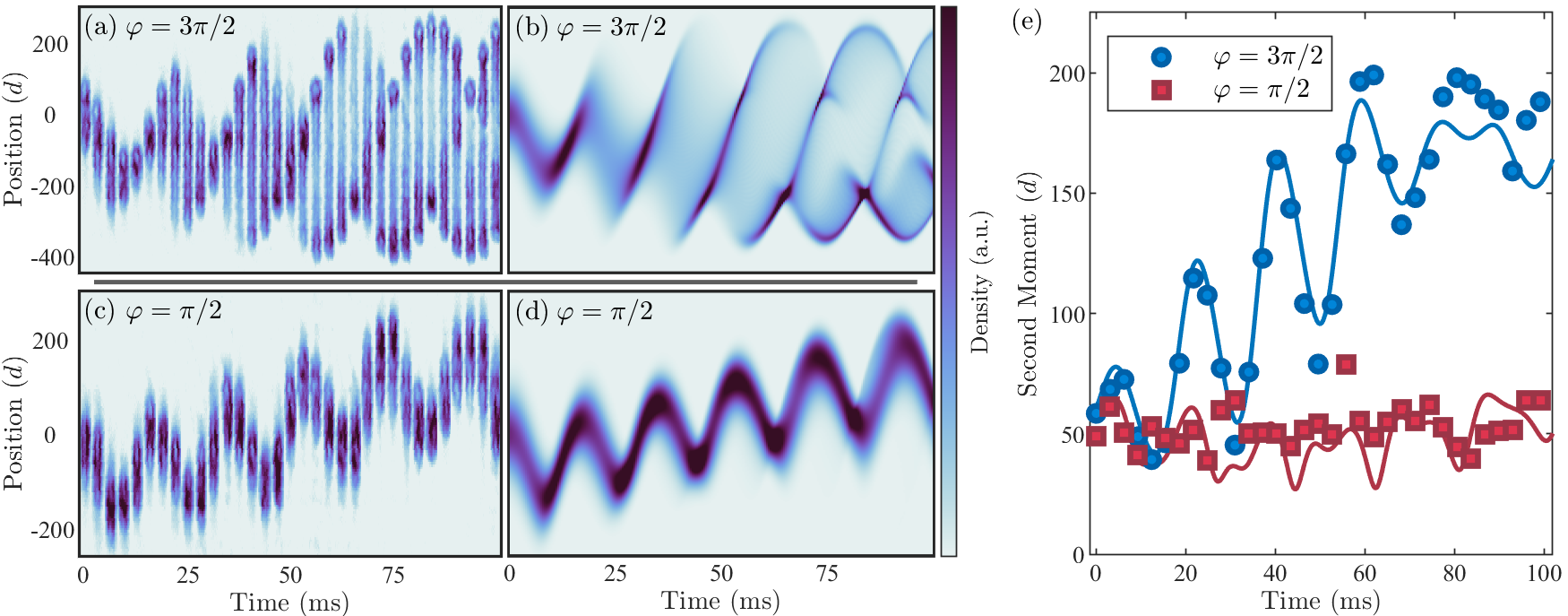}
    \caption{Phase-dependent spatial dynamics as a probe of the stroboscopic Poincar\'e map. (a) and (c): Time-sequence of  absorption images of an atomic ensemble subjected to drive phases of $\varphi=3\pi/2$ (a) and $\varphi=\pi/2$ (c). These two phases are predicted to give rise to topologically distinct Poincar\'e orbits with qualitatively different transport dynamics, as shown in Fig.~3c. 
    (b) and (d): Numerical simulations of the 1D density evolution under the same conditions as (a) and (c).
    The simulated density is averaged over independent, Gaussian-weighted 1D trials at varying lattice depths corresponding to transverse variation of beam intensity. (e) The second moment of the density distribution is plotted versus time for initial drive phases $\varphi = 3\pi/2$ (circles) and $\pi/2$ (squares).  Solid lines show simulated second moment evolution, accurately capturing the asymmetric enhancement and suppression of curvature-induced monopole modes.}
    \label{fig:secondmoment}
\end{figure*}

The breakdown of the theory based on a constant effective electric field motivates the search for a more complete theoretical description of driven Bloch dynamics in inhomogeneous fields.  As a key result of this work, we show that the qualitatively different dynamics observed arise from a rich underlying inhomogeneity-induced phase space structure which exhibits a topological transition in the character of stroboscopic Poincar\'e orbits. To see this, we use a Floquet map formalism, analyzing the phase-space trajectories stroboscopically with respect to the drive. Fig. \ref{fig:phasemap}a shows the calculated Poincar\'e map for a spatially uniform force and nonzero drive detuning. Super-Bloch-like oscillatory behavior is evident: the detuning generates a uniform quasimomentum shift for every point in phase space, and mediates the evolving phase in the sinusoidal arguments of Eq. \ref{eq:transport}, leading to an oscillatory profile of position versus quasimomentum. The sign of the quasimomentum shift is determined by whether the drive is red or blue-detuned, and the constant nature of the detuning ensures that the Floquet trajectories all wrap the phase space cylinder. On resonance, a transition occurs where all trajectories unwrap as the quasimomentum shift vanishes, and each point undergoes a vertical shift in the position direction determined by Eq.~\ref{eq:meanvg}; this is accompanied by the emergence of two fixed lines in the map for quasimomenta corresponding to initial drive phases of $\pi/2$ and $3\pi/2$.  The inset of Fig.~\ref{fig:phasemap}a shows the map for a resonant drive with starting points at only one initial position, revealing the invariant quasimomenta.

In the presence of force inhomogeneity, the fixed lines mentioned above become fixpoints where the drive meets the (position-dependent) Wannier-Stark resonance condition, yielding a strikingly different stroboscopic Poincar\'e map, as shown in Fig. \ref{fig:phasemap}b. The  fixpoints near $k=-0.5$ and $0.5 k_L$ are center-like and saddle-like respectively. Here super-Bloch-like transport breaks down and the system admits a new class of motion not present in Fig. \ref{fig:phasemap}a, namely regular cyclic orbits about the $k=-0.5 k_L$ fixed point. In the stroboscopic map, these orbits have a distinct topology as closed loops which do not wrap the Brillouin zone cylinder;  this emerges due to the possibility of the now implicitly time-dependent detuning changing its cycle averaged sign, something not possible in the homogeneous force case. A topologically distinct class of super-Bloch-like trajectories wrapping the Brillouin zone are observed at positions sufficiently far away from the resonance point. 

The Floquet map serves as a powerful intuitive tool for understanding and predicting the results of inhomogeneity in the effective electric field. As shown in \ref{fig:phasemap}c, the experimental $\varphi=\pi/2$ condition is represented by an ensemble which starts near the stable fixpoint, and the observed DC transport represents a partial cycle of the circulatory behavior in which the position spread of the ensemble is not significantly changed, approximating the motion of a rigid body. A drive phase of $\varphi=3\pi/2$, in contrast, corresponds to an ensemble starting near the saddle-like fixpoint. In this case the Floquet map dynamics predict rapid divergence along the unstable axis of the fixpoint, with the ensemble stretching in phase space and splitting up among orbits confined at the positional extremes. This corresponds to a maximal violation of rigid-body-like dynamics, exhibiting oscillations and growth in higher moments of the spatial distribution.

Our experimental observations confirm these predictions of the stroboscopic Poincar\'e map. Fig. \ref{fig:secondmoment}a and \ref{fig:secondmoment}c compare experimental image sequences of BEC evolution for drive phases of $\varphi=3\pi/2$ and $\varphi=\pi/2$ respectively. In the $\varphi=3\pi/2$ data, an initially localized distribution is observed to rapidly spread, eventually delocalizing over 600 lattice sites in a highly non-normal distribution. Here, the drive acts to amplify the curvature-induced breathing mode, eventually splitting the cloud largely into two regions of higher density near the edge of the distribution. Note that these dynamics correspond in detail to the predicted $\varphi=3\pi/2$ evolution illustrated in Fig~\ref{fig:phasemap}c. In contrast, the data for a drive phase $\varphi=\pi/2$, shifted by exactly $\pi$, reveals surprisingly stable wavepacket transport given the presence of both significant force inhomogeneity and a strong drive. In this case, the drive serves the dual role of preserving the wavefunction spatial mode and inducing transport. Both cases exhibit good agreement with numerical simulations of squared-wavefunction evolution shown in Fig. \ref{fig:secondmoment}b and \ref{fig:secondmoment}d. The match to theory is quantitative as well as qualitative: Fig. \ref{fig:secondmoment}e compares the evolution of the second moment of the position distribution to numerical theory for both drive phases. For $\varphi = 3\pi/2$, the second moment oscillates and grows rapidly before saturating at nearly 4 times the initial value. This is in stark contrast to the $\varphi=\pi/2$ case, in which the data display little to no variation of the second moment over the entire interval. 

In conclusion, we have shown that the combination of an inhomogeneous force with periodic modulation drives rich new dynamical behaviors beyond those of a pure Bloch or super-Bloch oscillator. Good agreement with numerical calculations supports our interpretation of the inhomogeneity-induced dynamics as arising from a fundamental change in the phase space structure of the Hamiltonian giving rise to distinct topological classes of Poincar\'e orbits.  These results point the way to a general protocol for controlling transport and density evolution with lattice amplitude modulation even in uncontrolled force environments. Potential future applications of these techniques include the generation of spatially squeezed states, new models for solid-state high-harmonic generation, and control elements for continuously-trapped atom interferometry. Since force metrology is an important use of atomic Bloch oscillations, these results have direct relevance for applications in which the force environment is both uncontrolled and inhomogeneous. Inclusion of static or modulated interatomic interactions is an exciting possible direction for future work~\cite{haqueBOs,modinteractions}, as is the effect of quasiperiodic or multiple-frequency driving. Such a platform would be well-suited for exploring the correspondence between the breakdown of classical orbit regularity and non-ergodic many-body dynamics. 

\begin{acknowledgements}
The authors thank Max Prichard for experimental assistance and Erich Mueller for useful discussions. DW acknowledges support from the National Science Foundation (CAREER 1555313), the Army Research Office (PECASE W911NF1410154 and MURI W911NF1710323), and the University of California's Multicampus Research Programs and Initiatives (MRP-19-601445). DW and RS acknowledge support from the UCSB NSF Quantum Foundry through Q-AMASE-i program award DMR-1906325.
\end{acknowledgements}


\bibliographystyle{apsrev4-2}

\end{document}